\documentclass[letter,twocolumn]{jpsj3}
\usepackage{bm}% bold math

\newcommand{\mbra}[1]{\left\{#1\right\}}

\title{Analisis of Shot Noise at Finite Temperatures in FQH Edge States}

\author{
\name{Eiki \surname{Iyoda}}\thanks{E-mail address: iyoda@issp.u-tokyo.ac.jp}, 
and \name{Tatsuya \surname{Fujii}}
}

\inst{
Institute for Solid State Physics, University of Tokyo, Kashiwa, Chiba 277-8581, Japan
}

\abst{
We investigate shot noise at {\it finite temperatures} 
induced by the quasi-particle tunneling between 
fractional quantum Hall (FQH) edge states. 
The resulting Fano factor has 
the peak structure at a certain bias voltage. 
Such a structure indicates that quasi-particles 
are weakly {\it glued} due to thermal fluctuation. 
We show that the effect makes it possible to 
probe the difference of statistics 
between $\nu=1/5,{\ }2/5$ FQH states where 
quasi-particles have the same unit charge.
Finally we propose a way to 
indirectly obtain statistical angle in hierarchical FQH 
states.}

\kword{fractional quantum Hall effect, Non-equilibrium Kubo formula, shot noise}

\begin{document}
\maketitle

The fractional quantum Hall (FQH) effect occurs 
in the two dimensional electron system 
subject to a strong magnetic field. 
The Hall conductance 
exhibits plateaus at filling fractions\cite{Tsui1982}. 
What is most surprising is that the phenomena is 
explained by existence of 
the quasi-particle with a fractional charge $e^*$
\cite{Laughlin1983}. 
The FQH state becomes an incompressible fluid with 
an energy gap. A confining geometry in low disorder samples 
makes the edges gapless modes, which carries 
fractionally charged particles. 
Thus direct observation of $e^*$ at edge states 
has been a stimulating problem. 

As the pioneering work Kane and Fisher proposed 
the setup to measure shot noise associated with 
the tunneling between counter-propagating edge states 
narrowed at a quantum point contact (QPC)~\cite{Kane1994}. 
It was shown that 
the Poisson noise at zero temperature 
is proportional to $e^*$ 
which arises from the granularity of charge. 
Inspired by the result, $e^*=e/3$ of the Laughlin quasi-particle 
was directly observed~\cite{Saminadayar1997,Picciotto1997}. 
The applicability has enhanced further 
studies of 
$e^*=e/5$ in a hierarchical FQH state with 
$\nu=2/5$\cite{Reznikov1999}.
The well-established scheme is applied to various filling factors, for example 
$\nu=5/2$\cite{Dolev2008} and  $\nu=2/3$\cite{Bid2009}. 

It is also predicted that 
the quasi-particle obeys fractional statistics
\cite{Wilczek1982}, 
characterized by a statistical angle $\theta$ 
upon an adiabatic exchange process of two quasi-particles. 
In two spatial dimensions, a statistical angle is 
allowed to take an intermediate one between $\theta=0$ 
for boson and $\theta=\pi$ for 
fermion, namely $\it anyons$. 
The issue to observe $\theta$ 
has also become an intriguing question. 

To detect $\theta$, 
on the basis of quantum interference,
it was proposed to use a kind of 
Aharanov-Bohm (AB) effect 
in equilibrium
\cite{Kivelson1990,Jain1993,Chamon1997,Kane2003,Rosenow2007}. 
The recent studies succeeded to detect 
the relative statistics
\cite{Camino2005}.
%\cite{Camino2005,Camino2007}.
Other attempts were made 
in Ref.[\citen{Isakov1999}], 
and Refs.[\citen{Safi2001,Vishveshwara2003}] 
where cross-correlation in three edge states was studied 
for the Laughlin states at zero temperature. 
Kim extended their works 
into hierarchical FQH states at finite frequency and temperature,
 and then discussed statistics for $\nu=1/5,{\ }2/5$\cite{Kim2005}.
It is well known that  
quasi-particles in $\nu=1/5,{\ }2/5$ FQH states 
have the same charge, but obey different statistics. 
Shot noise measurement in the low-temperature limit
succeeded to confirm the former 
\cite{Reznikov1999}, but failed to do the latter. 
Ref.[\citen{Kim2005}] gave an idea to 
the problem, but not realized in experiments. 

In this paper, we revert back to 
the standard set up
\cite{Kane1994,Saminadayar1997,
Picciotto1997,Reznikov1999,Dolev2008,Bid2009}, 
and discuss shot noise in
two edge states 
instead of three ones. 
What is most important point is 
that finite temperature effects are considered 
on the basis of our recent work~\cite{Fujii2007}. 
It is shown that the approach enables us to detect the difference of statistics 
in $\nu=1/5,{\ }2/5$ FQH states. 
Finally we discuss a method to determine statistical angle itself in hierarchical FQH states.

First of all, let us start with our formalism: 
shot noise at $\it finite{\ }temperatures$ 
based on the nonequilibrium Kubo formula 
in mesoscopic systems~\cite{Fujii2007}. 
It is well known that the usual Kubo formula 
determines linear conductance. 
In contrast, we derived 
a relation to differential conductance $G$ 
under finite bias voltages.
Thus this formula 
was called as the nonequilibrium Kubo formula. 
Then, it was also proposed to define shot noise at any temperature $S_h$ 
as the following formula: 
\begin{align}
\label{NELKubo2}
S_h
\equiv 
-\left<\left\{
\delta I_B,e^*(\delta N_L-\delta N_R
\right\}\right>,
\end{align}
where $\delta A\equiv A-\left<A\right>$. 
$I_B$ and $N_{L,R}$ are 
the backscattering current and number of 
quasi-particles in left/right reservoirs. 
It was proved that $S_h$ in eq.(\ref{NELKubo2}) has
several aspects expected as shot noise.
(i) In a non-interacting system, $S_h$
directly gives the Landauer-type shot noise at finite temperatures.
(ii) At zero temperature, 
$S_h$ is agreement with the standard 
shot noise: current noise $S$ at $T=0$.
(iii) In the linear response regime, 
$S_h=0$ and eq.(\ref{NELKubo1}) reproduces the Nyquist-Johnson relation.
As a result $S_h$ is qualified as shot noise at finite temperatures. 
Actually using $S_h$ it was successful to 
study shot noise of the Kondo effect in a quantum dot~\cite{Fujii2010,Yamauchi}. 

The nonequilibrium Kubo formula also satisfies 
the relation: 
\begin{align}
\label{NELKubo1}
S_h=S-4k_BTG,
\end{align}
where $S$ is current noise,
$G=\partial_V I_B$ is differential conductance,
$k_B$ is Boltzmann constant, 
and $T$ is temperature in reservoirs.
This relation shows us what variations of thermal noise should be subtracted from
current noise to define shot noise at finite temperatures in experiments.
Our framework gives a prospective 
way to study shot noise at finite temperatures: 
$S_h$ in eq.(\ref{NELKubo2}) is 
directly calculated, 
and its prediction is examined through 
$S-4k_{\rm B}TG$ in experiments. 
In the following the approach is applied to edge states.

Let us consider the quasi-particle tunneling through 
the QPC set at $x=0$ between edge states\cite{Kane1994,Wen1995,Martin2005}.
The gapless edge states are described by 
chiral Tomonaga-Luttinger liquids. 
We begin with Laughlin states 
with filling fractions $\nu=1/(2n+1)$ for simplicity. 
The Hamiltonian is given by right/left going edge 
modes $H_R, H_L$ and the tunneling part $H_B$: 
$H=H_R+H_L+H_B$, 
\begin{align}
H_{R,L}
&=
\frac{v_F}{\pi}\int^{\infty}_{-\infty}
dx
\left(
\frac{
\partial \phi_{R,L}(x)}
{
\partial x
}
\right)^2,
\\
H_B
&=
t_B
\psi_R^\dag(0)\psi_L(0)+h.c.,
\end{align}
where we put Planck constant $\hbar$ one,
$v_F$ is the Fermi velocity,
$t_B$ is the tunneling amplitude of quasi-particles,
$\phi_{R,L}(x)$ are chiral boson fields,
and $\psi_{R,L}(x)$ represent the vertex operators 
in $c=1$ CFT as
\begin{align}
\psi_{R,L}(x)=
\frac{1}{\sqrt{2\pi}} e^{\pm i k_F x} :e^{\mp i \sqrt{\nu}\phi_{R,L}(x)}:. 
\end{align}

The quasi-particle hopping with an unit charge $e^*$ 
generates backscattering current:    
\begin{align}
\hat{I}_B
\equiv
ie^* 
\left(
e^{i\omega_0t}
t_B
\psi^\dag_R(0)
\psi_L(0)
-
e^{-i\omega_0t}
t_B^*
\psi^\dag_L(0)
\psi_R(0)
\right). 
\end{align}
Following the discussion of a gauge transformation
\cite{Martin2005}, 
source-drain bias voltage $V$ is incorporated 
into phase factor $\omega_0=e^* V$.   
The backscattering current and current noise 
are obtained as
%\begin{align}
%I_B
%&=
%\nonumber
%\frac{1}{2}
%\sum_\alpha
%\left<
%T_K
%\left\{
%\hat{I}_B(t^\alpha)
%\exp
%\left(
%-i\int_Kdt_1H_B(t_1)
%\right)
%\right\}
%\right>,
%\\
%S
%&=
%\nonumber
%\sum_\alpha
%\left<
%T_K
%\left\{
%\delta \hat{I}_B(t^\alpha)
%\delta \hat{I}_B(t^{\bar{\alpha}})
%\exp
%\left(
%-i\int_Kdt_1H_B(t_1)
%\right)
%\right\}
%\right>,
%\end{align}
\begin{align}
\nonumber
I_B
&=
\left<\hat{I}_B(t)\right>,
\qquad
S
=
\int dt^\prime
\left<
\mbra{\hat{I}_B(t),\hat{I}_B(t^\prime)}
\right>.
\end{align}
%\begin{align}
%\nonumber
%I_B
%&=
%\left<I_B(t)\right>,
%\qquad
%S
%=
%\int dt^\prime \left<
%\mbra{I_B(t),I_B(t^\prime)}
%\right>.
%\end{align}
These quantities can be calculated on the basis of Schwinger-Keldysh formalism.

Conventionally information on a fractional charge is extracted 
from Poisson noise. Thus, when the quasi-particle weakly 
transmits through the QPC, it is sufficient to calculate 
these transport quantities 
up to the lowest order ${\cal O}(t_B^2)$. 
These expressions are rewritten into Landauer forms: 
\begin{align}
\label{Current}
I_B
&=
{e^*}
\int 
\frac{d\omega}{2\pi}
T(\omega)(f_L-f_R),
\\
S
&=
\nonumber
{e^*}^2
\int 
\frac{d\omega}{\pi}
T(\omega)(f_L(1-f_L)+f_R(1-f_R))\\
&+
\label{NoisePower}
{e^*}^2
\int 
\frac{d\omega}{\pi}
T(\omega)(f_L-f_R)^2.
\end{align}
The transmission probability is 
characterized by the $t$-matrix $t(\omega)\equiv \pi t_B \rho(\omega)$: 
\begin{align}
\label{Tt}
T
\left(\omega\right)
& \equiv
t^*
\left(
\omega-\omega_0/2
\right)
t
\left(
\omega+\omega_0/2
\right), 
\\
\nonumber
t(\omega)
&
=\frac{t_B}{v_F} \cosh\left(\frac{\beta\omega}{2}\right)
\left(\frac{2\pi}{\beta v_F}\right)^{\nu-1}
\frac{|\Gamma\left(\frac{\nu}{2}+i\frac{\beta\omega}{2\pi}\right)|^2}
{\pi\Gamma(\nu)}.
\end{align}
Here $\rho(\omega) \equiv 
-\mathrm{Im}G^r\left(\omega\right)/\pi$ is 
the density of states (DOS) at $x=0$. 
The quasi-particle Green function is defined by 
\begin{align}
G^{\alpha\beta}_{R,L}(t)
& \equiv -i\left< T_K
\psi_{R,L}(t^{\alpha})
\psi_{R,L}^{\dagger}(0^{\beta})\right>, \nonumber\\
G_{R,L}^{\pm\mp}(t)
&=
\frac{\pm i}{2\pi}
\left( \frac{i\frac{\pi}{\beta}}
{\sinh\frac{\pi}{\beta}(t\pm i\epsilon)}\right)^\nu,
\label{qG}
\end{align}
where $\epsilon$ is an infinitesimal positive number
where $\alpha, \beta=\pm$ represents a branch of the Keldysh contour
and the retarded component becomes $\mathrm{Im}G^r=i(G^{+-}-G^{-+})/2$.

In the $\nu=1$ IQH state or non-interacting edge states, 
$T(\omega)$ in eq.(\ref{Tt}) leads to a constant $|t_B|^2$. 
Within our approximation,
only the transmission probability is renormalized by Coulomb interaction
through the DOS.
In contrast the Fermi distribution function 
is unrenormalized as 
$f_{L,R}\equiv\frac{1}{1+\exp(\beta (\omega \pm \omega_0/2))}$.

Seemingly, the second line in eq.(\ref{NoisePower}) 
might be interpreted as shot noise in view of 
a Landauer-formula sense. 
However, shot noise formula in eq.(\ref{NELKubo2}) 
modifies the naive prediction. 
Following the same approximation, eq.(\ref{NELKubo2}) 
can be calculated, 
and rewritten into a Landauer-like form: 
\begin{align}
\label{eqD1}
S_h
&=S_L+\delta S_L,
\\
S_L &=
\nonumber
{e^*}^2 \int 
\frac{d\omega}{\pi}
T(\omega)(f_L-f_R)^2, 
\\
\delta S_L &=
\label{shl}
{e^*}^2 \int 
\frac{d\omega}{\pi}
T(\omega)(f_L-f_R)(y_L-y_R),
\end{align}
where $S_L$ represents a Landauer-type shot noise and
$\delta S_L$ does the correction term. Here,
\begin{align}
y(\omega)
&\equiv
\frac{1}{2}\tanh\left(\frac{\beta\omega}{2}\right)
-\frac{1}{\pi}\mathrm{Im}
\left[\Psi\left(\frac{\nu}{2}+i\frac{\beta\omega}{2\pi}
\right)\right],
\nonumber
\\
y_{L,R}
&\equiv y\left(\omega\pm\omega_0/2\right),
\label{y}
\end{align}
where $\Psi(z)$ is the digamma function. 
In case of $\nu=1$, 
because
$\delta S_L=0$
and $T(\omega)=|t_B|^2$,
it is exemplified that $S_h$ 
is equivalent to the Landauer-type shot noise $S_L$
for $\nu=1$
at finite temperatures. 
Thus the correction term $\delta S_L$ plays an essential 
role in FQH states. 

Let us discuss the feature in view of the nonequilibrium Kubo formula.
$G=\partial_V I_B$ is calculated using eq.(\ref{Current}), 
and then the resulting $G$ and $S$ are substituted 
into $S-4k_BTG$. 
Therefore, we confirm that the result 
is identical with $S_h$ in eq.(\ref{eqD1}), 
so that the nonequilibrium 
Kubo formula eq.(\ref{NELKubo1}) is satisfied. 
In the context it is found that 
$\delta S_L$ corresponds to the $V$-derivative of $T(\omega)$. 

\begin{figure}[tbp]
\begin{center}
\includegraphics[width=0.62 \columnwidth]{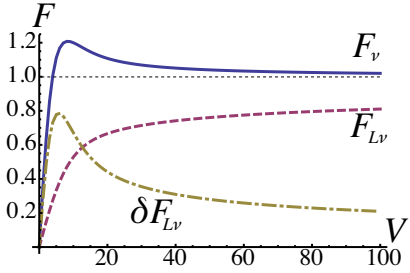}
\caption{
Bias dependence of $F_{\nu}$ (solid line), 
$F_{L\nu}$ (dashed line) and 
$\delta F_{L\nu}$ (dashed-dotted line) 
for $\nu=1/3$ at $\beta=1$}
\label{Peak1}
\end{center}
\end{figure}

To proceed a further discussion, we introduce 
the following Fano factor: 
\begin{align}
F_\nu\equiv\frac{S_h}{2e^*I_B}.
\label{FanoFactor1}
\end{align}
The different point compared to a standard Fano factor is 
to be normalized by an unit charge $e^*$. 
In the low-temperature/high-bias limit, 
$S_h/2I_B$ converges to $e^*$, as discussed later. 
With the normalization factor at zero temperature, 
it enables us to focus on thermal fluctuation of 
shot noise. 
Here in Fig.\ref{Peak1} $F_\nu$ for $\nu=1/3$ at a fixed 
inverse temperature $\beta=1$ is compared to $F_{L\nu}$ and 
$\delta F_{L\nu}$ defined by 
%
%\begin{align}
%F_{L\nu}\equiv \frac{S_L}{2e^*I_B},{\ \ }
%\delta F_{L\nu}\equiv \frac{\delta S_L}{2e^*I_B}. 
%\end{align}
$F_{L\nu}\equiv \frac{S_L}{2e^*I_B},{\ \ }
\delta F_{L\nu}\equiv \frac{\delta S_L}{2e^*I_B}$. 
$F_{L\nu}$ 
monotonously changes, on the other hand $\delta F_{L\nu}$ 
exhibits non-monotonousness 
with increasing bias voltage. 
The total Fano factor $F_{\nu}$ which is the sum of them
converges to $1$ in the high-bias limit. 
The fact represents that the charge of the quasiparticle is $e^*$ 
because $F_{\nu}$ is normalized by the unit charge. 
In contrast, we find the peak structure 
at a finite bias which originates from the correction 
noise $\delta F_{L\nu}$. 
It turns out that the enhancement from the unit charge occurs. 
Therefore, transmitted quasi-particles tend to 
come together due to thermal fluctuation. 

The peak structure is a sign for carried charges 
to bunch induced by thermal fluctuation. 
Therefore we call the effect "{\it thermal bunching}".
Note that this 
"{\it thermal bunching}" is different from the bunching 
which originates from statistics of quasi-particles 
\cite{Kim2005}. 

\begin{figure}[tbp]
\begin{center}
\includegraphics[width=0.9 \columnwidth]{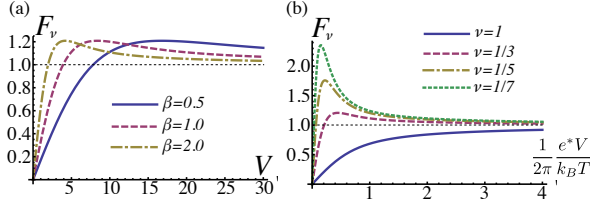}
\caption{(a) $F_\nu$ for $\beta=0.5,1.0,2.0$ at $\nu=1/3$ 
(solid line, dashed line, dash-dotted line);
(b) $F_\nu$ for $\nu=1,1/3,1/5,1/7$ at $\beta=1$ 
(solid line, dashed line, dash-dotted line, dotted line)}
\label{Peak2}
\end{center}
\end{figure}

Fig.\ref{Peak2}(a) shows 
$F_\nu$ at a fixed $\nu=1/3$ 
for several temperatures $\beta$. 
As lowering temperature, 
a peak position moves to a lower bias. 
What is universal nature is 
that the peak height is unchanged. 
On the other hand  in Fig.\ref{Peak2}(b) 
$F_\nu$ at a fixed $\beta=1$ is drawn 
for some parameters $\nu$. 
The peak structure does not appear in the $\nu=1$ IQH state. 
The peak height becomes larger for smaller $\nu$, 
namely, stronger magnetic field or Coulomb 
interaction. 
%We find that 
%the peak height diverges in the limit of $\nu \rightarrow 0$ 
%as, 
%%
%\begin{align}
%\lim_{\nu \rightarrow +0} F_\nu |_{\rm max} 
%\rightarrow \frac{1}{\pi \nu}.
%\end{align}
%%
%%
%\begin{figure}[htbp]
%\begin{tabular}{cc}
%\begin{minipage}{0.5\hsize}
%\begin{center}
%\includegraphics[width=0.9 \columnwidth]{fig2}
%\caption{$F_\nu$ for $\beta=0.5,1.0,2.0$ at $\nu=1/3$ 
%(solid line, dashed line, dash-dotted line)}
%\label{Peak2}
%\end{center}
%\end{minipage}
%\begin{minipage}{0.5\hsize}
%\begin{center}
%\includegraphics[width=0.96 \columnwidth]{fig3}
%\caption{$F_\nu$ for $\nu=1,1/3,1/5,1/7$ at $\beta=1$ 
%(solid line, dashed line, dash-dotted line, dotted line)}
%\label{Peak2}
%\end{center}
%\end{minipage}
%\end{tabular}
%\end{figure} 
%\begin{figure}[htbp]
%\begin{tabular}{cc}
%\begin{minipage}{0.5\hsize}
%\begin{center}
%\includegraphics[width=0.85 \columnwidth]{fig2a}
%\end{center}
%\end{minipage}
%\begin{minipage}{0.5\hsize}
%\begin{center}
%\includegraphics[width=0.96 \columnwidth]{fig2b}
%\end{center}
%\end{minipage}
%\end{tabular}
%\label{Peak2}
%\caption{(a) $F_\nu$ for $\beta=0.5,1.0,2.0$ at $\nu=1/3$ 
%(solid line, dashed line, dash-dotted line);
%(b) $F_\nu$ for $\nu=1,1/3,1/5,1/7$ at $\beta=1$ 
%(solid line, dashed line, dash-dotted line, dotted line)}
%\end{figure} 

Up to now, our discussion has been restricted to 
Laughlin states with $\nu=1/(2n+1)$. 
As said in the introduction,
we would like to consider statistics for $\nu=1/5,{\ }2/5$.
Here thus we extend our discussion into hierarchical FQH states 
(ex.$\nu=2/5,3/7,2/9,\cdots$).
Those states are characterized by 
filling fraction, unit charge and statistical angle: 
\begin{align}
\begin{array}{clllll}
\nu=\dfrac{p}{2np+1}& e^*=e\dfrac{1}{2np+1}& \theta=
\pi\dfrac{2n(p-1)+1}{2np+1}
\end{array}
\label{etn}
\end{align}
which are originally defined through the $K$-matrix\cite{Wen1995}. 
Our formalism developed above is described by 
the quasi-particle Green function at $x=0$. 
This treatment can be also justified when multiple tunneling processes
can be neglected(discuss later).
Thus the extension changes the exponent $\nu$ 
to $\alpha$ in eq.(\ref{qG}): 
\begin{align}
\alpha
=\dfrac{1}{p(2np+1)}.
\label{a}
\end{align}

Furthermore,
the current eq.(\ref{Current}), current noise eq.(\ref{NoisePower}) 
and shot noise eq.(\ref{eqD1}) 
have been expressed as a frequency-integral form, 
in accordance with 
the concept of Landauer formula. 
Concerning the current and current noise, integrated results 
were derived for Laughlin states\cite{Martin2005}.
%in Ref.[\cite{Martin2005}]. 
According to the same idea, shot noise can be also calculated. 
The result is straightforwardly extended into 
hierarchical FQH states, 
and thus the Fano factor is governed by a scaling 
function: 
\begin{align}
\label{FanoFactor2}
F_\alpha 
=
\frac{2}{\pi}
\mathrm{Im}
\left[
\Psi
	\left(
	\alpha+i\frac{1}{2\pi} 
        \left(\frac{e^* V}{k_BT}\right)
	\right)
\right].
\end{align}
This function is characterized by exponent $\alpha$.
If we define Fano factor as a ratio between total current noise $S$ and backscattering current $I_B$,
it only gives $\coth(\omega_0\beta/2)$, which does not include $\alpha$.

Taking advantage of the scaling form, 
let us reexamine the peak structure of the Fano factor. 
We plot $\partial_V F_\alpha$ taking $\alpha$ as a continuous parameter. 
It is found that $\alpha<1/2$ is the sufficient condition 
for emergence of peak. 
It is easily shown that $\alpha=1/p(2np+1)$ is less than 
$1/2$, and thus in all type of hierarchical FQH states 
a peak structure develops.

As an experimentally relevant case, 
let us discuss statistics 
in FQH states with $\nu=1/5,{\ }2/5$. 
The quantities listed in eq.(\ref{etn}) 
are specifically obtained for these states: 
\begin{center}
\small
\begin{tabular}{lll||c}
%%\hline
$\nu=1/5${\ \ \ }& $e^*_{1/5}=e/5${\ \ \ }
& $\theta_{1/5}=\pi/5${\ \ \ }&{\ }$(n,p)=(2,1)$ \\
\hline
$\nu=2/5${\ \ \ }& $e^*_{2/5}=e/5{\ \ \ }$ 
& $\theta_{2/5}=3\pi/5${\ \ \ }&
${\ }(n,p)=(1,2)$ 
%%\hline
\label{tab}
\end{tabular}
\end{center}
As mentioned in the introduction, 
$e^*_{1/5}=e^*_{2/5}=e/5$
has been confirmed through 
shot noise measurement in the low-temperature limit\cite{Reznikov1999}. 
However, there still remain the problem on statistics. 

To address the issue we begin with generic 
relations among $e^*$,$\theta$,$\nu$ and $\alpha$ 
in eqs.(\ref{etn}) and (\ref{a}). 
Each of $e^*$,$\theta$,$\nu$ and $\alpha$ 
is determined by two 
integer parameters: $n$ and $p$. 
Thus the independent quantities become two of them, 
and others are given by them. 
Therefore 
the statistical angle discussed here is represented 
in all type of hierarchical FQH states as 
\begin{align}
\label{Theta}
\theta=\pi\left[ 1-\alpha
\left(\frac{e}{e^*}-1\right)\right]. 
\end{align}
The result yields one-on-one relations 
between $\alpha$ and $\theta$
using $e^*_{1/5}=e^*_{2/5}=e/5$: 
\begin{align}
\label{at1/52/5}
\alpha_{1/5}=\frac{1}{4}\left( 1-\frac{\theta_{1/5}}{\pi}\right),
{\ }
\alpha_{2/5}=\frac{1}{4}\left( 1-\frac{\theta_{2/5}}{\pi}\right).
\end{align}
In conclusion, in order to see the difference of statistical 
angles there is a way to discuss 
exponents $\alpha$. 

\begin{figure}
\begin{center}
\includegraphics[width=0.62 \columnwidth]{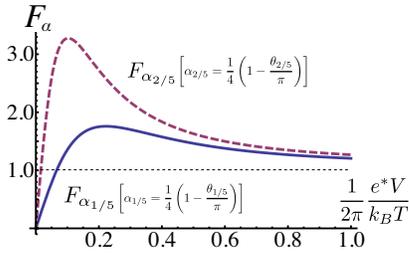}
\caption{
Fano factors of $\nu=1/5$ (Laughlin state) 
and $\nu=2/5$ (hierarchical FQH state).
Both quasi-particles have the same fractional charge $e/5$.
}
\label{Peak4}
\end{center}
\end{figure}

We 
substitute $\alpha_{1/5},{\ }\alpha_{2/5}$ 
in eq.(\ref{at1/52/5}) 
into eq.(\ref{FanoFactor2}) for $\nu=1/5,{\ }2/5$ 
respectively, and 
show the Fano factors for $\nu=1/5,{\ }2/5$ in Fig.(\ref{Peak4}). 
In the low-temperature/high-bias limit, 
both Fano factors converge to $1$. 
This is a generic feature of 
the Fano factor normalized by unit charge $e^*$. 
In the present case for $\nu=1/5,{\ }2/5$, 
even if the normalization by $e$ is considered, 
the limiting values are equal: 
$e_{1/5}^*/e=e_{2/5}^*/e=1/5$. 
It turns out that 
shot noise in the low-temperature/high-bias limit 
cannot distinguish statistics. 
What is striking is that
the Fano factors of $\nu=1/5,2/5$ have difference at finite temperature/bias.
%The Fano factor assigned to $\theta_{2/5}$ is 
%more enhanced than one to $\theta_{1/5}$ 
%at a finite temperature/bias. 
The Fano factor is determined by observable quantities thruogh eq.(\ref{NELKubo1}) and (\ref{FanoFactor1}). 
%Actually eq.(\ref{FanoFactor1}) proves that 
%$F_\alpha=(S-4k_BTG)/2e^*I_B$, 
%using the nonequilibrium Kubo formula $S_h=S-4k_BTG$. 
Analyzing its Fano factor, 
it is possible to
distinguish statistics of 
$\theta_{1/5}$ and $\theta_{2/5}$ in experiments. 

%%

%Here let us recall that 
%the Fano factor is determined by observable quantities. 
%Actually eq.(\ref{FanoFactor1}) proves that 
%$F_\alpha=(S-4k_BTG)/2e^*I_B$, 
%using the nonequilibrium Kubo formula $S_h=S-4k_BTG$. 
%Analyzing its Fano factor, 
%it is possible to
%distinguish statistics of 
%$\theta_{1/5}$ and $\theta_{2/5}$ in experiments. 

Having studied on specific statistics for $\nu=1/5,2/5$,
the last discussion  turns to issue to determine statistical angles of general hierarchical FQH states. 
As pointed out in eq.(\ref{Theta}), statistical angle 
$\theta$ is given by unit charge $e^*$ and exponent $\alpha$.
It is well established that 
an unit charge $e^*$ is determined through shot noise 
in the low-temperature limit. 
Note that exponent $\alpha$ is also available 
in the following. 
%As discussed above, the Fano factor 
%is a measurable quantity. 
By measuring the Fano factor $F_\alpha$, 
and then fitting the data to the scaling function in 
eq.(\ref{FanoFactor2}) as a parameter $\alpha$, 
one can infer the exponent $\alpha$.
Thus
we suggest an alternative procedure to determine $\theta$
by extracting $e^*$ and $\alpha$ from shot noise experiments.
The estimation of exponent $\alpha$ has been already reported by analyzing the power-law
dependence of tunneling current: $I\propto V^\alpha$\cite{Chang2003}.
Our approach makes it possible to obtain both $\alpha$ and $e^*$ in the shot noise measurement
with the scaling function eq.(\ref{FanoFactor2}) .

%We must discuss the limit of our treatment.
%Ferarro {\it et al.} shows multi-tunneling processes are not negligible 
%for hierarchical edge modes characterized by eq.(\ref{etn})
%at sufficiently lower temperature $T<T^*$\cite{Ferraro2008}. 
%The origin of the crossover temperature $T^*$ is dynamics of neutral edge modes.
%Our treatment is justified for $T>T^*$ and $T^*$ is evaluated as $50mK$ in an experiment\cite{Heiblum2003}/?/.

Finally we comment on closely-related works.
Ferarro et al.\cite{Ferraro2008} pointed out that the tunneling
is dominated by multiple particles at $T<T^*$,
in contrast the single particle at $T>T^*$.
The dynamics of neutral edge mode determines
the crossover temperature $T^*$, evaluated
as 50mK in the experiment\cite{Heiblum2003}.
Thus our treatment, which has focused on
the single-particle process, still stands
in the region of $T>T^*$.
As another work,
Isakov et al.\cite{Isakov1999} showed a monotonic-behavior of Fano factor in non-interacting particles with exclusion statistics.
Thus the interaction between particles is relevant for "thermal bunching".
%In our treatment,
%interactions between particles are relevant for the structure of Fano factor.

In summary the finite-temperature shot noise at FQH edge states 
has been studied on the basis of the nonequilibrium 
Kubo formula. 
The peak structure of Fano factor has been found 
at a bias voltage. 
We have named the phenomena "thermal bunching" 
because this is a sign for quasi-particles to weakly glue, 
mediated by thermal fluctuation. 
The phenomena has been determined by 
a scaling function characterized by 
an exponent of quasi-particle Green function. 
In $\nu=1/5,{\ }2/5$ FQH states, exponents have been 
given by only statistical angles. 
Detecting the discrepancy of Fano factors,
one can measure the difference of statistics. 
Finally we have proposed an indirect way to determine 
a statistical angle 
from exponent fitted to the scaling function 
and unit charge estimated 
at sufficiently low temperature within $T>T^*$
%%%%%%%%%%%%%%%%%%%%%%%%%%%%%%

The authors would like to thank T. Martin, 
J. Goryo, K. Kobayashi, N. Hatano, A. Furusaki, 
K. Ueda, and T. Kato for helpful discussions. 
E. I. acknowledge financial support by GCOE for Phys. Sci. Frontier, MEXT, Japan.
This research was partially supported by JSPS and MAE under the 
Japan-France Integrated Action Program (SAKURA).

%%%%%%%%%%%%%%%%%%%
%Bibliography

\end{document}